\documentclass[12pt]{article}

\usepackage{amsfonts,,amssymb}

\newcommand{\ra}{\rightarrow}
\newcommand{\Hom}{{\rm Hom}}

\newcommand{\CC}{{\mathbb C}}
\newcommand{\ZZ}{{\mathbb Z}}
\newcommand{\RR}{{\mathbb R}}

\newcommand{\PP}{{\mathbb P}}

\newcommand{\tw}{{\tilde w}}
\newcommand{\modtwo}{{\ {\rm mod}\ 2}}

\newcommand{\dis}{\sqcup}
\newcommand{\cA}{{\mathcal A}}
\newcommand{\cT}{{\mathcal T}}
\newcommand{\im}{{\rm im}}

\title{Symmetry Protected Topological Phases, Anomalies, and Cobordisms: Beyond Group Cohomology}

\author{Anton Kapustin \\ {\it California Institute of Technology}}

\begin{document}

\maketitle

\abstract{We propose that Symmetry Protected Topological Phases with a finite symmetry group $G$ are classified by cobordism groups of the classifying space of $G$. This provides an explanation for the recent discovery of bosonic SPT phases which do not fit into the group cohomology classification. We discuss the connection  of the cobordism classification of SPT phases to gauge and gravitational anomalies in various dimensions.}

\section{Introduction and summary}

An important problem in condensed matter theory is to understand equivalence classes of gapped phases of matter with a symmetry group $G$ and no long-range entanglement. The equivalence is understood in the sense of homotopy theory. Such equivalence classes are called Symmetry Protected Topological (SPT) phases. The classification of free fermionic SPT phases with or without translational invariance is well understood by now \cite{Kitaev}, but the situation with interacting systems is more complicated. We will be interested in the interacting case here and for simplicity will discuss only theories without fermions. We do not assume translational invariance, so the SPT phases we discuss are robust with respect to disorder.

It has been proposed that bosonic SPT phases in space-time dimension $d$ are classified by the degree-$d$ cohomology group of $BG_0$ with $U(1)$ coefficients \cite{GC}. Here $G_0$ is a finite symmetry group\footnote{A version suitable for compact Lie group symmetries has also been proposed in \cite{GC}, but we will limit ourselves to finite groups here.} which is purely internal, i.e. acts trivially on space-time, and $BG_0$ is its classifying space. The space $BG_0$ classifies $G_0$ bundles in the sense that isomorphism  classes of principal $G_0$ bundles on any space $X$ are in one-to-one correspondence with homotopy classes of maps from $X$ to $BG_0$. 
Cohomology of $BG_0$ is also known as group cohomology of $G_0$. 

In the case when the symmetry group $G$ involves time-reversing elements, it has been proposed that bosonic SPT phases are classified by elements of the degree-$d$ cohomology group of $BG$ with coefficients in a certain flat $U(1)$ bundle over $BG$ \cite{GC}. However, it was noticed later that certain 4d bosonic SPT phases with time-reversal symmetry do not fit into this classification \cite{VS1,VS2,VS3}. 

In this paper we attempt to refine the classification based on group cohomology. Related ideas have been considered in \cite{Wenanomalies}. We propose that in space-time dimension $d$ bosonic SPT phases with a finite internal symmetry group $G_0$ and vanishing thermal Hall response  are classified by the Pontryagin-dual of the torsion subgroup of the degree-$d$ oriented bordism group of $BG_0$. 

The oriented bordism group of a space $X$ is usually denoted $\Omega_{SO,d}(X)$. Its Pontryagin-dual is defined as 
$$\Hom(\Omega_{SO,d}(X),U(1)).$$ 
We will denote this group $\Omega_{SO}^d(X,U(1))$ and call it the oriented cobordism group of $X$ with $U(1)$ coefficients.\footnote{This terminology and notation are not standard. To motivate them, recall that the Pontryagin-dual of the integral homology group of a space $X$ is the cohomology of $X$ with $U(1)$ coefficients. On the other hand, the integral cohomology group of $X$ is related to the integral homology in a more complicated way, through the universal coefficient formula. The cobordism group of $X$ as usually defined is analogous to the integral cohomology group of $X$, while what we need here is an analogue of the cohomology with $U(1)$ coefficients.} Similarly, we define 
$$
\Omega_{SO}^d(X,\RR)=\Hom(\Omega_{SO,d}(X),\RR).
$$ 
The map $e: \RR\ra U(1)$ which sends $x$ to $\exp(2\pi i x)$ induces a map 
$$
e: \Hom(\Omega_{SO,d}(X),\RR) \ra \Hom(\Omega_{SO,d}(X),U(1)).
$$
The image of this map consists of elements of $\Omega_{SO}^d(X,U(1))$ which vanish on the torsion subgroup of $\Omega_{SO,d}(X)$. Therefore the Pontryagin-dual of the torsion subgroup of $\Omega_{SO,d}(X)$ can be identified with the quotient $\Omega_{SO}^d(X,U(1))/\im\ e$. Thus we propose that bosonic SPT phases in $d$ dimensions with a finite internal symmetry group $G_0$ and vanishing thermal Hall response are classified by $\Omega^d_{SO}(BG_0,U(1))/\im\ e$.

More generally, if some elements of the symmetry group $G$ are time-reversing, we are given a homomorphism $\rho: G\ra \ZZ_2$. Its kernel $G_0$ is the group of internal symmetries. We propose that  in space-time dimension $d$ bosonic SPT phases with symmetry $G$ and vanishing thermal Hall response are classified by the degree-$d$ oriented cobordism group of $BG$ with twisted coefficients, with the twist determined by $\rho$ (see below for a detailed explanation). This twisted cobordism group will be denoted $\Omega^d_{SO}(BG,U(1)^\rho)$.  In this case one does not need to quotient by the cobordism group with real coefficients because the twisted bordism group is pure torsion. 

For any space $X$ with a principal $\ZZ_2$ bundle $\rho$ we have homomorphisms from  $H^d(X,U(1))$ to $\Omega_{SO}^d(X)$ and from $H^d(X,U(1)^\rho)$ to $\Omega_{SO}^d(X,U(1)^\rho)$,  where $U(1)^\rho$ is a $U(1)$ principal bundles on $X$ associated to the $\ZZ_2$ principal bundle $\rho$ on $X$ via the obvious action of $\ZZ_2$ on $U(1)$. Specializing to the case $X=BG_0$ or $X=BG$, we get a map from the group cohomology classification of SPT phases \cite{GC} to the cobordism classification. In general, this map is neither injective nor surjective. That is, there exist SPT phases which appear to be nontrivial from the group cohomology point of view, but are trivial from the cobordism point of view. There also exist SPT phases which are nontrivial from our point of view but are not captured by the group cohomology classification. The latter phenomenon occurs starting with $d=4$, while the former occurs only for $d>6$. Thus for $d\leq 6$ the cobordism classification is indeed a refinement of the group cohomology classification. The relation between the two classification schemes is discussed further in the concluding section.

SPT phases in dimension $d$ are related to 't Hooft anomalies in dimensions $d-1$ \cite{FT,Wenanomalies,KapThornanomalies}. These 't Hooft anomalies are the ones which can be canceled by anomaly inflow from $d$ dimensions (i.e. they are ambiguous phases in the partition function of the gauged system). Thus our results also provide a classification of such 't Hooft anomalies. We will consider a few concrete examples of systems with 't Hooft anomalies below. 

Even if the internal symmetry group $G_0$ is trivial, our proposed classification scheme gives a nontrivial set of SPT phases labeled by elements of $\Omega^d_{SO}(pt,U(1))/\Omega^d_{SO}(pt,\RR)$ (if no time reversal symmetry is present) or $\Omega_O^d(pt,U(1))$ (if time reversal symmetry is present). This is because the cobordism classification takes into account gravitational anomalies of the boundary phase. Equivalently, while the bulk of the SPT phase has a unique ground-state on a spatial slice of any topology (this is one possible interpretation of the ``no long-range entanglement'' condition), the partition function on a general space-time may be a complex number with absolute value $1$. 

Let us note a few special cases. For trivial $G_0$ and $G=\ZZ_2^T$ (i.e. SPT phases with only time-reversal symmetry), we have \cite{Thom}
\begin{eqnarray}
\Omega_O^1(pt,U(1)) & = &\Omega_O^3(pt,U(1))=0,\nonumber \\
\Omega_O^2(pt,U(1)) & = &\Omega_O^5(pt,U(1))=\ZZ_2,\nonumber \\
\Omega_O^4(pt,U(1)) & = &\ZZ_2\times\ZZ_2,\nonumber 
\end{eqnarray}
This agrees with the group cohomology classification up to dimension $3$. In dimension $4$ we find two extra nontrivial bosonic SPT phases with $\ZZ_2^T$ symmetry. Presumably, they can be identified with the new phases found in \cite{VS1} and further studied in \cite{VS2,VS3}. We also find a new nontrivial SPT phase in $d=5$ where group cohomology predicts no nontrivial SPT phases. This phase also arises in the oriented case, i.e. with no symmetry at all. Indeed, $\Omega^5_{SO}(pt,U(1))=\ZZ_2$ and $\Omega^5_{SO}(pt,\RR)=0$, so their quotient is $\ZZ_2$. An analogous phase does not exist in theories with fermions since $\Omega_{Spin}^5(pt,U(1))=0$ \cite{spincobordism}.

\section{Bordisms and cobordisms}

This section explains the necessary mathematical background following Atiyah \cite{Atiyah}.
Let $X$ be a topological space. A degree-$d$ oriented bordism to $X$ is a closed oriented $d$-manifold $M$ together with a continuous map $f:M\ra X$. There is an equivalence relation on bordisms: $(M_1,f_1)\sim (M_2,f_2)$ if there exists a $d+1$-dimensional compact oriented manifold $N$ with boundary $M_1\dis\bar M_2$ and a map $g:N\ra X$ which reduces to $f_1$ and $f_2$ on the two components of $\partial N$. The set of equivalence classes of oriented bordisms to $X$ forms an abelian group called the oriented bordism group denoted $\Omega_{SO, d}(X)$.  The group operation arises from the disjoint union operation on $d$-manifolds, and the negative of $(M,f)$ can be represented by $(\bar M,f)$, where $\bar M$ is the orientation reversal of $M$. The Pontryagin dual of $\Omega_{SO, d}(X)$ will be called the degree-$d$ oriented cobordism group of $X$ and will be denoted $\Omega^d_{SO}(X)$. There is a map $\Omega_{SO,d}(X)\ra H_d(X,\ZZ)$ which sends $(M,f)$ to $f_*[M]$, where $[M]\in H_d(M,\ZZ)$ is the fundamental homology class of $M$ (the distinguished generator of $H_d(M,\ZZ)$). This map is a homomorphism of abelian groups. Since the dual of $H_d(M,\ZZ)$ is $H^d(M,U(1))$, the dual map can be thought of as a homomorphism $H^d(X,U(1))\ra \Omega_{SO}^d(X)$. In general, these maps are neither injective nor surjective. 

Recall that a manifold $M$, whether oriented or not, carries a canonical principal $\ZZ_2$ bundle $\xi_M$ called the orientation bundle of $M$. The holonomy of the associated $\ZZ_2$-connection assigns $-1$ to all closed loops which reverse orientation and assigns $+1$ to all closed loops which do not. A closed manifold $M$, whether oriented or not, has the fundamental class living in $H_d(M,\ZZ^T)$, where $\ZZ^T$ is a local system with fiber $\ZZ$ twisted by the orientation bundle. Its Pontryagin-dual is $H^d(M,U(1)^T)$, where $U(1)^T$ is a local system over $M$ with fiber $U(1)$ twisted by the orientation bundle.

Let $X$ be a topological space together with a principal $\ZZ_2$ bundle $\rho$ over it. A degree-$d$ twisted oriented bordism to $(X,\rho)$ is a closed, but unoriented and perhaps unorientable, $d$-manifold $M$ together with a map $f:M\ra X$ and an isomorphism $f^*\rho\simeq \xi_M$. There is an obvious way to define an equivalence relation on twisted oriented bordisms. The set of equivalence classes of twisted oriented bordisms to $(X,\rho)$ is an abelian group denoted $\Omega_{SO,d}(X,\rho)$. Its Pontryagin dual will be called the group of twisted oriented cobordisms of $X$ with coefficients in $U(1)^\rho$ and will be denoted $\Omega^d_{SO}(X,U(1)^\rho)$.  Note that if $\rho$ is a trivial bundle, a twisted oriented bordism is the same as an oriented bordism to $X$. Note also that there is a map $\Omega_{SO,d}(X,\rho)\ra H_d(X,\ZZ^\rho)$ which sends $(M,f)$ to $f_*[M]$. This map is a homomorphism of abelian groups. Its dual is a homomorphism from $H^d(X,U(1)^\rho)$ to $\Omega^d_{SO}(X,U(1)^\rho)$.

Oriented bordism groups of a point have been determined by Wall \cite{Wall}. For other spaces the answer is more difficult to compute. Note however that in the case $X=B\ZZ_2$ and $\rho$ the universal $\ZZ_2$ bundle $E\ZZ_2$ over $B\ZZ_2$, a twisted oriented bordism to $(B\ZZ_2,E\ZZ_2)$ is the same as an unoriented bordism to a point. Hence in this special case we have $\Omega_{SO,d}(X,\rho)=\Omega_{O,d}(pt)$. The latter group has been computed by R. Thom \cite{Thom}.

\section{Thermal Hall response}

Our basic assumption is that a gapped state of matter with short-range interactions can be put on a curved space-time of arbitrary topology, and that at long distances the partition function can be computed using field theory.  This is likely to apply to a wide range of gapped systems, not just SPT phases. At short distances a system is usually defined on a regular lattice, with short-range interactions. However, if we allow for disorder, then dislocations in the lattice are possible, and more general triangulations also become possible. If the system admits a Euclidean lattice formulation, this applies both to space and time directions. 

Consider now an SPT phase $\alpha$ in dimension $d$ with a finite internal symmetry group $G_0$. With the above assumption, we can compute its  partition function for any closed oriented $d$-manifold $M$ equipped with a $G_0$-connection $A$. This partition function can be thought of as the value of $\exp(2\pi i S^\alpha_M(A))$, where $S^\alpha_M$ is the effective action for $A$. This action is gauge-invariant and is an integral of a local Lagrangian. The partition function is a pure phase because by assumption the ground state of the SPT phase is unique on a spatial slice of any geometry.  Because of CPT theorem the effective action satisfies an important property $S^\alpha_{\bar M}(A)=-S^\alpha_M(A)$. It is also additive under disjoint union, while the partition function is multiplicative. We will denote by $\bar\alpha$ the orientation-reversal of the SPT phase $\alpha$. By definition, $S^{\bar\alpha}_M(A)=S^\alpha_{\bar M}(A)$, therefore $S^{\bar\alpha}_{\bar M}(A)=-S^{\alpha}_M(A)$. 

In general the effective action may contain local geometric terms. They can be of two kinds: the terms which depend only on the topology of $M$ and the terms which also depend on the geometry of $M$. In this section we discuss the latter.\footnote{I am grateful to Alexei Kitaev for pointing out that such geometric terms are allowed.}

Consider the functional derivative of $S^\alpha_M$ with respect to the metric on $M$. If this derivative does not vanish identically, this means that there is a nontrivial vacuum expectation value of the stress-energy tensor when the system is placed into a nontrivial geometric background and/or coupled to a background gauge field. Such SPT phases have a nontrivial thermal Hall response. In the case when $G_0$ is trivial, the corresponding action must depend only on the metric and be odd under orientation-reversal. This implies that it must be a gravitational Chern-Simons term. Such terms exist in dimensions $d$ of the form $4n-1$. More generally, one can have also mixed terms which involve the gauge field $A$ as well. 

We can simplify the problem by focusing on SPT phases with vanishing thermal Hall response. The effective action then does not depend on the metric and is purely topological.

Limiting ourselves to systems with vanishing thermal Hall response is not a very serious limitation. Indeed, any two SPT phases with the same thermal Hall response differ by an SPT phase with vanishing thermal Hall response. Thus the classification of general SPT phases is equivalent to the classification of SPT phases with vanishing thermal Hall response plus the classification of gravitational Chern-Simons terms (including the mixed ones). The latter problem is fairly straightforward.

\section{Cobordisms and SPT phases}

Having disposed of the geometric terms in the effective action, we now ask how it can depend on the topology of $M$. It is instructive to look at the case of trivial $G_0$ first. Then the action depends only on the topology of $M$. Since it is also local, it must be an integral of products of characteristic class of $M$. In the oriented case (i.e. without time-reversal symmetry) there are three kinds of such classes: the Pontryagin classes, with exist in dimensions divisible by $4$, the Stiefel-Whitney classes, which exist in all dimensions, and the Euler class, which exists in top dimension (but vanishes if the dimension is odd). The Euler class is ruled out, because it is odd under orientation-reversal, and thus its integral over $M$ will be even, while the action is supposed to be odd. Thus $S^\alpha_M$ must be a linear combination of integrals of products of Pontryagin and Stiefel-Whitney classes over $M$. These integrals are called Pontryagin and Stiefel-Whitney numbers, respectively.

It is well-known that Pontryagin and Stiefel-Whitney numbers of $M$ depend only on the oriented bordism class of $M$ \cite{Thom}. Thus the effective action can be viewed as a map from $\Omega_{d,SO}(pt)$ to $U(1)$. Since it is additive under disjoint union of manifolds, it is actually a group homomorphism. If $\Omega_{d,SO}(pt)$ contains a free part (this happens for $d$ divisible by $4$), then the corresponding action contains continuous theta-parameters. Varying such parameters does not change the SPT phase, hence we should should identify SPT phases which differ only the values of these theta-parameters. Equivalently, one can say that an SPT phase is characterized by an element of
$\Hom(\Omega_{d,SO}(pt),U(1))$ modulo the image of $\Hom(\Omega_{d,SO}(pt),\RR)$.

The case when the only symmetry is the time-reversal symmetry is very similar. In that case $M$ is unoriented and carries no further data. The only relevant characteristic classes in this case are Stiefel-Whitney classes. Since Stiefel-Whitney numbers depend only on the unoriented bordism class of $M$, we conclude that the effective action is a homomorphism from $\Omega_{d,O}(pt)$ to $U(1)$. Since all elements in $\Omega_{d,O}(pt)$ have order $2$, in this case the effective action does not contain any continuous parameters, and we conclude that bosonic SPT phases are labeled by elements of $\Omega^d_O(pt,U(1))$. 

On the basis of these two examples we propose that in general $S^\alpha_M$ is cobordism-invariant. More precisely, In the case when the symmetry group $G_0$ is internal (does not involve time-reversal), the gauge field $A$ can be thought of as a map $\cA:M\ra BG_0$. Thus the pair $(M,\cA)$ can be thought of as an oriented bordism to $BG_0$. We propose that the effective action depends only on the equivalence class of this bordism.
The partition function $\exp(2\pi i S^\alpha_M(A))$ then can be thought of as a map from $\Omega_{d,SO}(BG_0)$ to $U(1)$. This map is multiplicative under the disjoint union of bordisms and therefore is a homomorphism of groups. Identifying actions  which differ only by the value of continuous parameters is equivalent to taking the quotient of $\Hom(\Omega_{d,SO}(BG_0),U(1))$ by  the image of $\Hom(\Omega_{d,SO}(BG_0),\RR)$.

Consider now a more general case when the symmetry group $G$ involves some time-reversing elements. This is described by a homomorphism $\rho:G\ra\ZZ_2$ whose kernel $G_0$ consists of internal symmetries. Since reversing time reverses orientation of space-time, for a nontrivial $\rho$ the SPT phases $\alpha$ and $\bar\alpha$ are isomorphic. Therefore $2S^\alpha_M(A)=0$ for all $M$ and $A$, i.e. $S^\alpha_M(A)=0$ or $1/2$. Thus all nontrivial SPT phases have order $2$. Further, since $\alpha$ is identified with $\bar\alpha$, one can define the model on an unorientable manifold $M$. The $G$-gauge field is partially determined by the geometry of $M$. Indeed, consider a loop $\gamma$ on $M$ which reverses orientation. The holonomy of the $G$-connection around $\gamma$ should lie in the time-reversing part of $G$, i.e. $\rho$ should map it to the nontrivial element of $\ZZ_2$. On the other hand, if $\gamma$ is orientation-preserving, the holonomy around $\gamma$ should lie in $G_0=\ker\rho$. One can describe such a $G$-connection as follows. A general $G$-connection on $M$ defines a map $\cA: M\ra BG$. Given $\rho$, we have a canonical $\ZZ_2$ principal bundle over $BG$ obtained by applying the homomorphism $\rho$ to the fibers of the universal $G$-bundle over $BG$. (Equivalently, $\rho$ induces a map $BG\ra B\ZZ_2$, and one can use this map to pull back the universal $\ZZ_2$ bundle over $B\ZZ_2$ to $BG$).  We will also call it $\rho$. The constraint on the $G$-connection is that $\cA^*\rho\simeq\xi_M$. In other words, $(M,\cA)$ defines a twisted oriented bordism to $(BG,\rho)$. 

If we assume that $S^\alpha_M(A)$ is cobordism-invariant, then the effective action becomes a homomorphism from $\Omega_{d,SO}(BG,\rho)$ to $U(1)$. If $\rho$ is nontrivial, all elements in $\Omega_{d,SO}(BG,\rho)$ have order $2$, hence the action does not contain continuous parameters. Thus we conclude that SPT phases in this case can be labeled by elements of $\Omega^d_{SO}(BG,U(1)^\rho)$. 

\section{Bosonic SPT phases protected by time-reversal symmetry}

In this section we compare the cobordism classification of bosonic SPT phases with time-reversal symmetry with the group cohomology classification.
As explained above, such SPT phases are classified by the group $\Omega_{SO}^d(B\ZZ_2,E\ZZ_2)$. By definition, a twisted oriented bordism to $(B\ZZ_2,E\ZZ_2)$ is a map $f$ from a closed $d$-manifold $M$ to $B\ZZ_2$ such that the pull-back of the universal $\ZZ_2$-bundle is $\xi_M$, the orientation bundle of $M$. Since the map to $BG$ is determined up to homotopy by the pull-back of the universal bundle, this means that up to homotopy $f$ is determined by the orientation bundle of $M$. Hence the set of equivalence classes of twisted oriented bordisms in this case is the same as the set of equivalence classes of unoriented bordisms to a point. That is, bosonic SPT phases with only time-reversal symmetry are classified by the unoriented cobordism group $\Omega^d_O(pt,U(1))$. 

The graded group  $\Omega^*_O(pt,U(1))=\oplus_d\ \Omega^d_O(pt,U(1))$ (or rather its dual $\Omega_{O,*}(pt)$) has been computed by R. Thom \cite{Thom} and has a simple structure. $\Omega_{O,*}(pt)$ is actually a graded ring, and can be identified with the ring of polynomials with $\ZZ_2$ coefficients in an infinite number of variables $x_j$ for all $j>0$ which are not of the form $2^i-1$ for some natural $i$. The degree of the variable $x_j$ is $j$. Thus the unoriented bordism ring has generators in degree $2,4,5,6,8,\ldots$. In low dimensions the unoriented bordism groups are
$$
\Omega_{1,O}(pt)=\Omega_{3,O}(pt)=0,\quad \Omega_{2,O}(pt)=\Omega_{5,O}(pt)=\ZZ_2,\quad \Omega_{4,O}(pt)=\ZZ_2\times\ZZ_2.
$$

One can write down explicitly the topological actions corresponding to all these SPT phases as integrals of polynomials of the Stiefel-Whitney classes. These are special cohomology classes $w_j\in H^j(M,\ZZ_2)$ which exist for all $j$ in the range $0<j\leq d=\dim M$. The lowest ones have a transparent geometric meaning. For example, $w_1$ is a connection on the orientation bundle of $M$, while $w_2$ is the obstruction to having a spin structure on $M$. The integral of the top class $w_d(M)$ is the Euler characteristic modulo $2$. There are relations between products of Stiefel-Whitney classes whose form depends on $d$. Thom's theorem says that the unoriented bordism class of $M$ is determined by the Stiefel-Whitney numbers of $M$ (i.e. integrals of polynomials of Stiefel-Whitney classes over $M$). Taking into account the relations between Stiefel-Whitney classes leads to the above result for the unoriented  cobordism groups of a point. 

For example, for $d=2$ we have a unique topological action given by the integral of $w_1^2$, which is equal to the integral of $w_2$, which in turn is equal to the Euler characteristic modulo $2$. Thus there is a unique nontrivial bosonic SPT phase in $d=2$  characterized by the fact that the partition function on any unorientable closed 2-manifold is $-1$, while on any orientable closed 2-manifold it is $1$. This agrees with the group cohomology classification \cite{ChenGuWen,FK}.

For $d=3$ no non-trivial bosonic SPT phase with $\ZZ_2^T$ symmetry is possible, because $w_1 w_2=w_3=w_1^3=0$ for all closed 3-manifolds, or equivalently, because any closed 3-manifold is a boundary of some compact 4-manifold.  This also agrees with the group cohomology classification \cite{GC}.

For $d=4$ there are two independent topological actions, since $w_3 w_1=w_2 w_1^2=0$, and $w_4+w_2^2+w_1^4=0$. A possible choice of generators for $\Omega_{SO,d}^4(pt,U(1))$ is $w_1^4$ and $w_2^2$. Thus we expect three nontrivial bosonic SPT phases in $d=4$, in agreement with \cite{VS2}. Note that group cohomology classification sees only one of these three phases. We can figure out which action corresponds to this special SPT phase by noting that the group cohomology approach is based on classifying actions which depend only on  the gauge field on $M$, which in this case is the first Stiefel-Whitney class $w_1\in H^1(M,\ZZ_2)$. Thus the group cohomology approach detects the action
$$
S_M=\frac12 \int_M w_1^4,
$$
but is unable to see $w_2^2$. Note that on spin manifolds $w_2=0$, so the additional SPT phases are possible only in theories without fermions.

For $d=5$ the only non-trivial action is
$$
S_M=\frac12 \int_M w_2 w_3.
$$
One can show that all other polynomials of Stiefel-Whitney classes of total degree $5$ vanish, including $w_1^5$. Thus there is a unique nontrivial bosonic SPT phase in $d=5$ with $\ZZ_2^T$ symmetry. It is not detected by group cohomology, because $H^5(B\ZZ_2,U(1)^T)=0$. This phase actually persists even if time-reversal symmetry is broken, since $\Omega_{5,SO}(pt)=\ZZ_2$. 

\section{Bosonic time-reversal anomalies}

As remarked in the introduction, SPT phases in $d$ dimensions are related to 't Hooft anomalies in $d-1$ dimensions. For example, time-reversal-invariant SPT phases in dimension $d$ correspond to 't Hooft anomalies for time-reversal in dimension $d-1$. Gauging time-reversal means defining the theory on an unoriented manifold, so an 't Hooft anomaly for time-reversal means that a theory has a time-reversal symmetry on a flat space-time, but nevertheless cannot be consistently defined on an unoriented manifold. This can be regarded as a special case of global gravitational anomaly.

Let us consider a couple of examples. The case $d=2$ is somewhat degenerate, since there are no unorientable 1-manifolds. Still, one can say that a time-reversal symmetry for a 1d system cannot be gauged if all states are odd under time-reversal. On the other hand, if such a 1d system is a boundary of a 2d SPT phase, then time-reversal can be gauged, because a spatial slice of a 2d SPT phase always looks like a collection of intervals, and all 1d edges come in pairs. 

In $d=4$ the situation is more interesting. Consider an action 
$$
S=\frac12 \int_M w_1^4.
$$
As argued above, this action describes a 4d SPT phase which fits into the group cohomology classification. On a 4-manifold with a boundary this action is not ``gauge-invariant'' and needs to be coupled to an anomalous 3d theory to compensate for it. To see how this works, let us represent the class $w_1$ by an integral 1-cochain $\tw_1$ satisfying $\delta \tw_1=0 \modtwo$.  $\tw_1$ is not uniquely defined, we have ``gauge transformations''
\begin{equation}\label{wonegauge}
\tw_1\ra \tw_1+\delta h+2\alpha,
\end{equation}
where $h$ is an arbitrary integral 0-cochain and $\alpha$ is an arbitrary integral 1-cochain. These transformations preserve the cocycle condition $\delta \tw_1=0 \modtwo$. Under such transformations the action changes by a boundary term:
$$
S\mapsto S+\frac12 \int_{\partial M} h (\delta h)^3.
$$
Here we dropped terms which are integral, since $S$ is defined modulo integers. 

To cancel this boundary term, we need to place on the boundary $\partial M$ a 3d theory which has an 't Hooft anomaly for the time-reversal symmetry. It turns out a simple topological $\ZZ_2$ gauge theory can do the job. This theory has an action
$$
S_{3d}=\frac12 \int_{\partial M} a \delta b,
$$
where $a$ and $b$ are integral 1-cochains. They should be thought of as $\ZZ_2$ gauge fields because the action is invariant under
\begin{equation}\label{ztwogauge}
a\mapsto a+\delta f+2\alpha,\quad b\mapsto b+\delta g+2\beta,
\end{equation} 
where $\alpha$ and $\beta$ are integral 1-cochains and $f,g$ are integral 0-cochains.

The 3d action $S_{3d}$ is also invariant under time-reversal because its value is half-integral on any configuration and thus $\exp(2\pi i S_{3d})=\pm 1$ is invariant under complex conjugation. One can of course promote this model  to a well-defined theory on an unoriented 3-manifold by leaving the action as it is. 
But one can also do something more interesting.  Let us couple the model to $\tw_1$ by adding an extra term to the action
\begin{equation}\label{Sprime}
S'_{3d}=\frac12\int_{\partial M} \left(a\delta b+ (a+b) \tw_1^2\right). 
\end{equation}
The action is still half-integral, so the theory appears to be time-reversal invariant. However, $S'_{3d}$ transforms nontrivially under the ``gauge transformations'' (\ref{wonegauge}):
$$
S'_{3d}\ra S'_{3d}+\frac12 \int_{\partial M} (a+b) (\delta h)^2.
$$
We can try to rectify this by postulating a nontrivial transformation law for $a$ and $b$:
\begin{equation}\label{gaugeh}
a\mapsto a+h\delta h,\quad b\mapsto b+h\delta h.
\end{equation}
Note that this is distinct from the $\ZZ_2$ gauge transformations (\ref{ztwogauge}) since $h\delta h=\delta (h^2/2)$ is not a coboundary of an integral 0-cochain. Then the action $S'_{3d}$ transforms  as follows:
$$
S'_{3d}\mapsto S'_{3d}+\frac12 \int_{\partial M} h(\delta h)^3.
$$
This variation cancels the variation of the bulk topological action, as desired.

It has been argued that the surface of the SPT phase predicted by group cohomology can be  described by a $\ZZ_2$ gauge theory with a projective action of time-reversal  symmetry \cite{VS2}. Specifically, the generator $\cT$ of $\ZZ_2^T$ satisfies $\cT^2=-1$ when acting on electric and magnetic quasiparticles. The effective action (\ref{Sprime}) describes precisely such a phase. To see this, note that in the TQFT language quasiparticles correspond to topological Wilson loop observables. In an ordinary $\ZZ_2$ gauge theory these observables are
$$\
W_a(\gamma)=\exp\left(\pi i \oint_\gamma a\right),\quad W_b(\gamma)=\exp\left(\pi i \oint_\gamma  b\right).
$$
They take values $\pm 1$ and depend only on the homotopy class of the loop $\gamma$. These loop observables correspond to electric and magnetic quasiparticles. 

On the other hand, in the theory defined by the action (\ref{Sprime}) the naive Wilson loops $W_a$ and $W_b$ are not topological observables because they are not invariant under the transformations (\ref{gaugeh}). However one can define the following modified observables:
$$
W_a(\gamma)=\exp\left(\pi i \oint_\gamma \left(a+\frac12 \tw_1\right)\right),\quad W_b(\gamma)=\exp\left(\pi i \oint_\gamma \left(b+\frac12 \tw_1\right)\right).
$$
They are topological because 
$$
\delta (a+\frac12 \tw_1)=\delta a+\tw_1^2 \modtwo, \quad \delta (a+\frac12 \tw_1)=\delta a+\tw_1^2\modtwo. 
$$
Note however that the new Wilson loops can now take values $\pm i$ if $\gamma$ is orientation-reversing. We conclude from this that electric and magnetic quasiparticles are eigenstates of $\cT$ with eigenvalues $\pm i$.  This agrees with the discussion in \cite{VS2}.

As our final example, let us consider the bosonic SPT phase in $d=5$.  The corresponding topological action is
$$
S=\frac12 \int_M w_2 w_3.
$$
One can check that all other Stiefel-Whitney numbers vanish. For a closed 5-manifold $M$ this action is invariant  under ``gauge transformations''
$$
w_2\mapsto w_2+\delta\alpha,\quad w_3\mapsto w_3+\delta\beta,
$$
where $\alpha$ is a 1-cochain with values in $\ZZ_2$ and $\beta$ is a 2-cochain with values in $\ZZ_2$. When $M$ has a nonempty boundary, the action changes by a boundary term:
$$
S\mapsto S+\frac12 \int_{\partial M} \left(\alpha w_3+\beta w_2+\alpha\delta\beta\right).
$$
Here we do not need to distinguish between $w_2$ and $w_3$ of $M$ and $\partial M$ because $w_2$ and $w_3$ of the normal bundle of $\partial M$ vanish for dimensional reasons.  Thus one needs to place on $\partial M$ a nontrivial $d=4$ theory which couples to $w_2$ and $w_3$. It is similar to the $d=3$ theory considered above. Namely, it is a $\ZZ_2$ topological gauge theory which in flat space can be described by an action
$$
S_{4d}=\frac12 \int_{\partial M} a\delta b,
$$ 
where $a$ is a 1-cochain with values in $\ZZ_2$, and $b$ is an 2-cochain with values in $\ZZ_2$. We couple it to $w_2$ and $w_3$ by modifying the action as follows:
$$
S'_{4d}=\frac12 \int_{\partial M} \left(a\delta b+ a w_3+b w_2\right)
$$
and postulating the following ``gauge transformations'' for $a$ and $b$:
$$
a\mapsto a+\alpha,\quad b\mapsto b+\beta.
$$
One can easily check that the variation of the boundary action $S'_{4d}$ cancels the variation of the bulk action.

\section{Concluding remarks}

It would be interesting to demonstrate directly that the new 4d bosonic SPT phase with $\ZZ_2^T$ symmetry proposed in \cite{VS1,VS2,VS3} can be described by a topological action based on Stiefel-Whitney classes $w_2^2$ or $w_4$. To this end it would be sufficient to compute the partition function on $\CC\PP^2$ and show that it is equal to $-1$ rather than $1$. Alternatively, one could try to show that the surface 3d theory proposed in \cite{VS1,VS2,VS3} has an 't Hooft anomaly for the time-reversal symmetry. 

An important issue is the relation between group cohomology and cobordism classifications of SPT phases. As discussed above, there is a natural map from the former to the latter (the dual map from bordisms to homology is known as the Thom homomorphism \cite{Thom}). We have seen above that the map is not onto already for $G=\ZZ_2^T$ and $d=4$, as well as $d=5$. That is, there exist bosonic SPT phases which are not captured by the group cohomology classification. From our point of view, this happens because group cohomology classification essentially treats $w_1$ (the first Stiefel-Whitney class) as a $\ZZ_2$ gauge field and ignores higher Stiefel-Whitney classes which can also enter the topological action $S^\alpha_M(A)$. For $d\leq 3$ this is not a great loss, since $w_2=w_1^2$ and $w_3=0$, while higher classes do not contribute. But starting with $d=4$ new SPT phases appear.

Going in the opposite direction, one may ask whether there exist bosonic SPT phases which are nontrivial from the point of view of group cohomology but  are trivial from the cobordism point of view. At first sight this appears unlikely, since a nontrivial class in $H^d(BG,U(1))$ corresponds to a nontrivial topological action $S_M^\alpha(A)$. Nevertheless, this might happen because the topology of smooth orientable manifolds is more constrained than the topology of general topological spaces. That is, there may exist nontrivial classes in $H^d(BG,U(1))$ which integrate to zero when pulled back to any closed 
oriented $d$-manifold. As shown in \cite{Thom}, this can happen only for $d>6$ (the example given in \cite{Thom} corresponds to $G=G_0=\ZZ_3\times\ZZ_3$ and $d=7$), so the issue is largely academic. That is, for $d\leq 6$ the cobordism classification of SPT phases is strictly finer than the group cohomology classification. 

While we discussed here bosonic theories, it should be straightforward to extend the classification to fermionic SPT phases by replacing (oriented)  cobordisms groups with (s)pin cobordism groups. It would be interesting to compare the resulting classification with the K-theory classification \cite{Kitaev}. Presumably the well-known relation between  K-theory and cobordism groups of a space plays an important role here.

I am would like to thank Ryan Thorngren for a collaboration on a related project, Dan Freed for drawing my attention to cobordism groups, and  John Morgan and Michael Hopkins for advice. I am especially grateful to Alexei Kitaev for pointing out a number of erroneous statements in the first version of the manuscript. This work was supported in part by the DOE grant  DE-FG02-92ER40701.

\end{document}